\pdfoutput=1
\documentclass[a4paper,11pt]{article}
\usepackage{pos}

\providecommand{\repositoryInformationSetup}{} 
\repositoryInformationSetup

\usepackage{xspace}
\usepackage{bbm}
\usepackage{amsmath}
\usepackage{mathtools}
\usepackage{dsfont}
\usepackage{braket}
\usepackage{upgreek}
\usepackage[T1]{fontenc}
\usepackage[boxruled,lined,commentsnumbered]{algorithm2e}
\usepackage[utf8]{inputenc}
\usepackage[UKenglish]{babel}
\usepackage{siunitx}
\usepackage{placeins}
\usepackage{bm}
\usepackage{tikz}
\usepackage{ifthen}
\usepackage{calculator}
\usepackage{tikz-feynman}
\usepackage{graphicx}
\usepackage{slashed}
\usepackage{xfrac}
\usepackage{changes}

\graphicspath{{data/},{figures/},{pics/}}

\newcommand{\repoURL}{https://github.com/evanberkowitz/hubbard-critical-coupling}

\newcommand{\Figref}[1]{Figure~\ref{fig:#1}\xspace}

\renewcommand{\Ref}[1]{Ref.~\cite{#1}}

\newcommand{\half}{\ensuremath{\frac{1}{2}} }                                   

\newcommand{\average}[1]{\ensuremath{\left\langle #1 \right\rangle}\xspace}

\newcommand{\erwartung}[1]{\ensuremath{\left\langle#1\right\rangle}}

\newcommand{\pdagger}{{\phantom{\dagger}}}

\newcommand{\meff}{\ensuremath{m_\text{eff}}\xspace}

\newcommand{\transpose}{\ensuremath{{}^{\top}}}
\newcommand{\adjoint}{\ensuremath{{}^{\dagger}}}

\newcommand{\spc}{single-particle correlator}

\let\builtinLaTeX\LaTeX
\def\LaTeX{\builtinLaTeX\xspace}

\newcommand{\grapheneColor}[2]{
	\foreach \x in {0,...,#1}
	{
		\MODULO{\x}{2}{\xmod} 
		\foreach \y in {0,...,#2}
		{
			\tikzset{yshift={\y*1.732cm+Mod(\x,2)*0.866cm}, xshift={\x*1.5cm}}
			\begin{scope}
				\draw[line width=2pt] (0,0)--(1,0);
				\ifthenelse{\xmod=0 \OR \y<#2 \AND \x>0}{\draw[line width=2pt] (0,0)--(-0.5, 0.866);}{}
				\ifthenelse{\xmod>0 \OR \y>0 \AND \x>0}{\draw[line width=2pt] (0,0)--(-0.5, -0.866);}{}
				\ifthenelse{\x=1 \AND \y=0}{\node (A1) at (0,0) {}; \node (A2) at (1.5, 0.866) {}; \node (A3) at (1.5, -0.866) {};}{}
				
				\fill[red] (0,0) circle (4pt);
				
				\fill[blue] (1,0) circle (4pt);
				\ifthenelse{\xmod=0 \OR \y<#2 \AND \x>0}{\fill[blue] (-0.5, 0.866) circle (4pt);}{}
				\ifthenelse{\y > 0 \OR \xmod > 0 \AND \x>0}{\fill[blue] (-0.5, -0.866) circle (4pt);}{}
			\end{scope}
}}}

\newcommand{\grapheneColorShift}[4]{
	\foreach \x in {0,...,#1}
	{
		\MODULO{\x}{2}{\xmod} 
		\foreach \y in {0,...,#2}
		{
			\tikzset{yshift={\y*1.732cm+Mod(\x,2)*0.866cm+#4}, xshift={\x*1.5cm+#3}}
			\begin{scope}
				\draw[line width=2pt] (0,0)--(1,0);
				\ifthenelse{\xmod=0 \OR \y<#2 \AND \x>0}{\draw[line width=2pt] (0,0)--(-0.5, 0.866);}{}
				\ifthenelse{\xmod>0 \OR \y>0 \AND \x>0}{\draw[line width=2pt] (0,0)--(-0.5, -0.866);}{}
				\ifthenelse{\x=1 \AND \y=0}{\node (A1) at (0,0) {}; \node (A2) at (1.5, 0.866) {}; \node (A3) at (1.5, -0.866) {};}{}
				
				\fill[red] (0,0) circle (4pt);
				
				\fill[blue] (1,0) circle (4pt);
				\ifthenelse{\xmod=0 \OR \y<#2 \AND \x>0}{\fill[blue] (-0.5, 0.866) circle (4pt);}{}
				\ifthenelse{\y > 0 \OR \xmod > 0 \AND \x>0}{\fill[blue] (-0.5, -0.866) circle (4pt);}{}
			\end{scope}
}}} 

\newcommand{\critU}{\ensuremath{\num{3.835(14)}}}

\newcommand{\critNu}{\ensuremath{\num{1.181(43)}}} 

\newcommand{\critExp}{\ensuremath{\num{0.898(37)}}}

\title{The Semimetal-Antiferromagnetic Mott Insulator Quantum Phase Transition of the Hubbard Model on the Honeycomb Lattice}

\author*[a]{Johann Ostmeyer}
\author[b,c]{Evan Berkowitz}
\author[c,d]{Stefan Krieg}
\author[c,e]{Timo A. L\"{a}hde}
\author[a,c,e]{Thomas Luu}
\author[a]{Carsten Urbach}

\affiliation[a]{
    Helmholtz-Institut f\"{u}r Strahlen- und Kernphysik,\\
    Rheinische Friedrich-Wilhelms-Universit\"{a}t Bonn, 53012 Bonn, Germany
}

\affiliation[b]{
	Maryland Center for Fundamental Physics,\\
	University of Maryland, College Park 20742, USA
}

\affiliation[c]{
	Institute for Advanced Simulation,\\
	Forschungszentrum J\"{u}lich, 54245 J\"{u}lich, Germany
}

\affiliation[d]{
	J\"{u}lich Supercomputing Center,\\
	Forschungszentrum J\"{u}lich, 54245 J\"{u}lich, Germany
}

\affiliation[e]{
    Institut f\"{u}r Kernphysik,\\
    Forschungszentrum J\"{u}lich, 54245 J\"{u}lich, Germany
}

\emailAdd{ostmeyer@hiskp.uni-bonn.de}
\emailAdd{e.berkowitz@fz-juelich.de}
\emailAdd{s.krieg@fz-juelich.de}
\emailAdd{t.laehde@fz-juelich.de}
\emailAdd{t.luu@fz-juelich.de}
\emailAdd{urbach@hiskp.uni-bonn.de}

\abstract{
	The Hubbard model on the honeycomb lattice undergoes a quantum phase transition from a semimetallic to a Mott insulating phase and from a disordered to an anti-ferromagnetically phase. We show that these transitions occur simultaneously and we calculate the critical coupling $U_c=\critU$ as well as the critical exponents $\nu=\critNu$ and $\beta=\critExp$ which are expected to fall into the $SU(2)$ Gross-Neveu universality class. For this we employ Hybrid Monte Carlo simulations, extrapolate the single particle gap and the spin structure factors to the thermodynamic and continuous time limits, and perform a data collapse fit. We also determine the zero temperature values of single particle gap and staggered magnetisation on both sides of the phase transition.
}

\FullConference{%
	The 38th International Symposium on Lattice Field Theory, LATTICE2021
	26th-30th July, 2021
	Zoom/Gather@Massachusetts Institute of Technology
}

\begin{document}

\maketitle

\allowdisplaybreaks[1]


\section{Introduction}

The Fermi-Hubbard model---a prototypical model of electrons hopping between lattice sites---has a rich phenomenology of strongly-correlated electrons, requiring nonperturbative 
treatment.
On a honeycomb lattice, where it provides a basis for studying electronic properties of carbon nanosystems like graphene and nanotubes, the Hubbard model is expected to exhibit a second-order quantum phase transition between a (weakly coupled) semi-metallic (SM) state and an antiferromagnetic Mott insulating (AFMI) state as a function of the electron-electron coupling.

It is noteworthy that a fully-controlled \emph{ab initio} characterization of the SM-AFMI transition has not yet appeared; Monte Carlo (MC) calculations have not provided unique, 
generally accepted values for the critical exponents~\cite{Otsuka:2015iba,semimetalmott}. Such discrepancies are primarily attributed to the adverse scaling of MC algorithms with spatial system
size $L$ and inverse temperature $\beta$. 
The resulting systematic error is magnified by an incomplete understanding of the extrapolation of operator expectation values to the thermodynamic and temporal continuum limits.

%

\begin{figure}[ht]
	\input{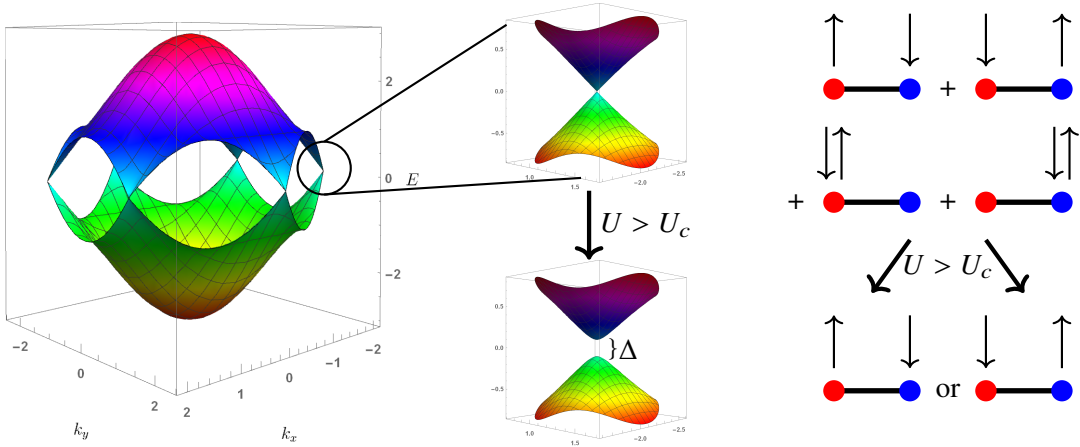}
	\caption{Left: The two energy bands (in multiples of the hopping $\kappa$) of the non-interacting Hubbard model as a function of the momentum $k$ normalised by the lattice spacing $a$. Center: Inset showing the Dirac cones. A band gap $\Delta$ separating the bands opens in the phase transition, once a critical coupling $U_c$ is surpassed. The bottom figure is only a qualitative visualisation, not the exact result. Right: The sublattice symmetry is broken at the same critical coupling and the disordered state (a superposition of all possibilities) transitions to an antiferromagnetic order. We show in the following (and in~\cite{semimetalmott,more_observables}) that the transitions happen simultaneously.}
	\label{fig:opening_gap}
\end{figure}

In this work we summarise the recent progress made, using Lattice Monte Carlo (LMC) techniques, in removing he systematic uncertainties which affect determinations of the critical exponents of
the SM-AFMI transition in the honeycomb Hubbard model~\cite{semimetalmott,more_observables}.
We present a unified, comprehensive, and systematically controlled treatment of both the single particle gap and the staggered magnetisation, the former being the electric and the latter the anti-ferromagnetic (AFM) order parameters of the phase transition. 
We thus confirm the AFM nature of the transition from first principles. The simultaneous opening of the gap and emergence of AFM order is illustrated in figure~\ref{fig:opening_gap}.
We find that the critical coupling $U_c/\kappa=\critU$ and 
the critical exponents---expected to be the exponents of the SU(2) Gross-Neveu, or chiral Heisenberg, universality class~\cite{heisenberg_gross_neveu}---to be $\nu=\critNu$ and $\upbeta=\critExp$.

We formulate the grand canonical Hubbard model at half filling in the particle-hole basis and without a bare staggered mass.
Its Hamiltonian reads
\begin{equation}
	H=-\kappa \sum_{\erwartung{x,y}}\left(p^\dagger_{x}p^\pdagger_{y}+h^\dagger_{x}h^\pdagger_{y}\right)+\frac{U}{2}\sum_{x}\rho_x\rho_x\,,\label{eqn_particle_hole_hamiltonian}
\end{equation}
where $p$ and $h$ are fermionic particle and hole annihilation operators, $\kappa$ is the hopping amplitude, $U$ the on-site interaction, and
\begin{equation}
	\rho_x = p^\dagger_x p_x -h^\dagger_x h_x,
	\label{eq:charge_operator}
\end{equation}
is the charge operator.
Using Hasenbusch-accelerated HMC with the BRS formulation~\cite{2018arXiv180407195K,Brower:2012zd} and a 
mixed time differencing~\cite{semimetalmott} which has favorable computational scaling properties, 
we generate ensembles of auxiliary field configurations for different linear spatial extents $L$ with a maximum of $L$=$102$ corresponding to 20,808 lattice sites, interaction strengths $U$ (with fixed hopping $\kappa$), inverse temperatures $\beta = 1/T$, and $N_t$ Trotter steps;\footnote{Throughout this work, we use an upright $\upbeta$ for the critical exponent and a slanted $\beta$ for the inverse temperature.}
See \Ref{semimetalmott} for full details.



\section{Observables}\label{sec:observables}

\subsection{The single particle gap}

\begin{figure}[t]
	\centering
	\includegraphics[width=.45\textwidth]{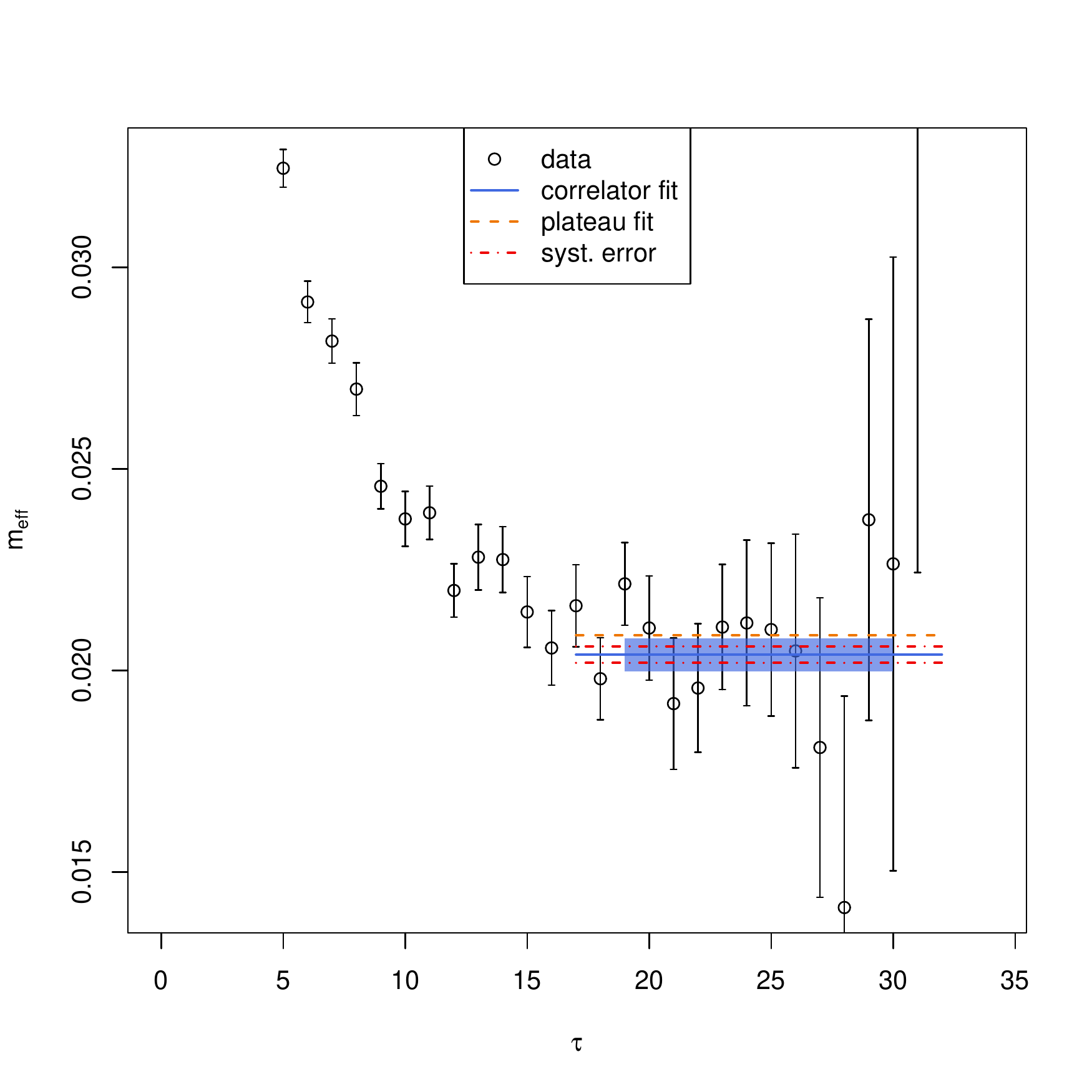}
	\includegraphics[width=.45\textwidth]{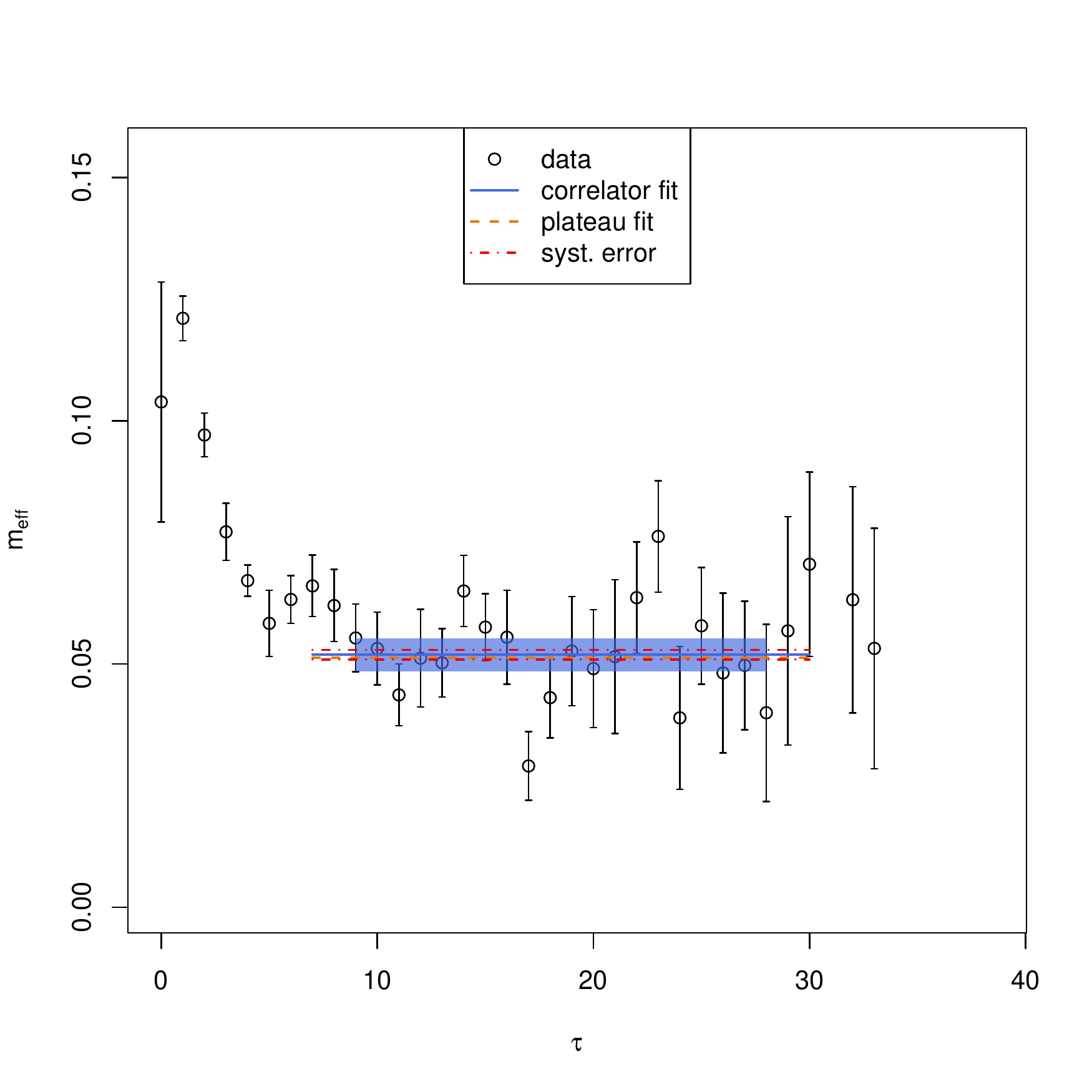}
	\caption{Examples of effective mass determinations from the {\spc}s, extracted from ensembles of auxiliary field configurations.
		The blue line with error band gives $\meff$ with statistical error, obtained from a hyperbolic cosine effective mass \eqref{eq:cosh fit}.
		The length of the blue band indicates the fitting region. For comparison, a constant fit to the effective local mass is 
		shown by the dashed orange line. The dot-dashed red line shows the estimation of the systematic error~\cite{semimetalmott}.
		Note that the red and orange lines have been extended outside of the fitting region, for clearer visibility.
		Left panel: $\kappa\beta=8$, $L=15$, $N_t=64$, $U/\kappa=\num{3.5}$. Right panel: $\kappa\beta=12$, $L=6$, $N_t=72$, $U/\kappa=\num{3.85}$.
		The timeslice index $\tau$ is integer-valued. The effective mass \meff is given in units of $\kappa$.
		\label{fig:plateau_fit}}
\end{figure}

The single particle gap measures the distance between the two energy bands which is closest at the Dirac momenta $K$ and $K'$. It can readily be obtained from the single particle correlator
\begin{align}
	C(t) \coloneqq \erwartung{p_{K,t}^\pdagger p\adjoint_{K,0}} \equiv \erwartung{p_{K',t}^\pdagger p\adjoint_{K',0}}
\end{align}
at said momenta in imaginary time by a fit of the effective mass $\meff$ to the functional form
\begin{align}
	C(\tau) = a \cosh\left(\meff\,\delta\left(\tau-N_t/2 \right)\right),
	\label{eq:cosh fit}
\end{align}
with $a$ and $\meff$ as fit parameters in a reasonable scaling region. Here $\tau=t/\delta$ has integer values and we symmetrised the correlator about $t=\beta/2$. For an example of such a fit see figure~\ref{fig:plateau_fit} and for more details see Ref.~\cite{semimetalmott}.
The single particle gap is simply given by
\begin{equation}
	\Delta = 2 \meff,
\end{equation}
for a specific value of $L$ and $\delta$.

\subsection{Spin structure factors}

At half-filling, we compute expectation values of one-point and two-point functions of bilinear local operators, 
the spins 
\begin{equation}
	S^i_{x} = \half (p^\dagger_x,\, (-1)^x h_x) \sigma^i (p_x,\, (-1)^x h^\dagger_x)\transpose
\end{equation}
and the charge~\eqref{eq:charge_operator}
where the $\sigma^i$ are Pauli matrices and $(-1)^x$ provides a 
minus sign depending on the triangular sublattice of the honeycomb to which the site $x$ belongs; the sign originates in the particle-hole transformation.
In this work we focus on the staggered sum of spins 
\begin{equation}
	S^i_{-} = \sum_x (-1)^x S^i_{x},
\end{equation}
which computes the difference between the sublattices, though the uniform sum of spins and the charge operators are readily available too and have been investigated in Ref.~\cite{more_observables}. We omit them here because they are no order parameters of the phase transition of interest.

Furthermore we define the correlator
\begin{equation}
	S^{ii}_{--} = \average{S^i_- S^i_-}.
\end{equation}
%
%
Because the dynamical exponent $z=1$~\cite{PhysRevLett.97.146401} we can obtain intensive order parameters at any finite temperature by taking the thermodynamic limit of these extensive quantities and dividing them by $\beta^2$, rather than the spatial volume $V=L^2$. 



\section{Analysis}\label{sec:analysis}

\begin{figure}[t]
	\centering
	\raisebox{0.016\height}{\includegraphics[width=.45\textwidth]{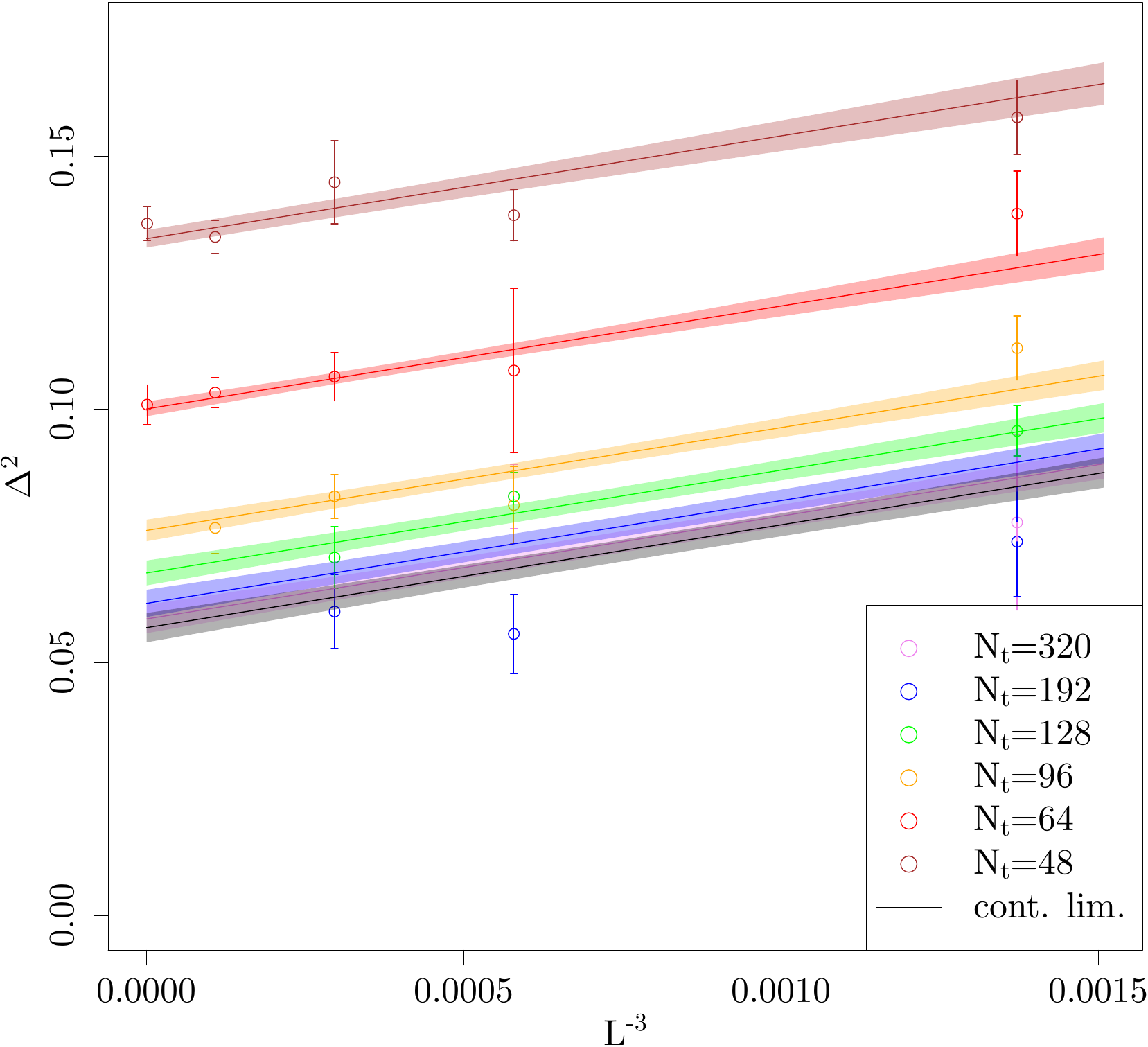}}
	\includegraphics[width=.45\textwidth]{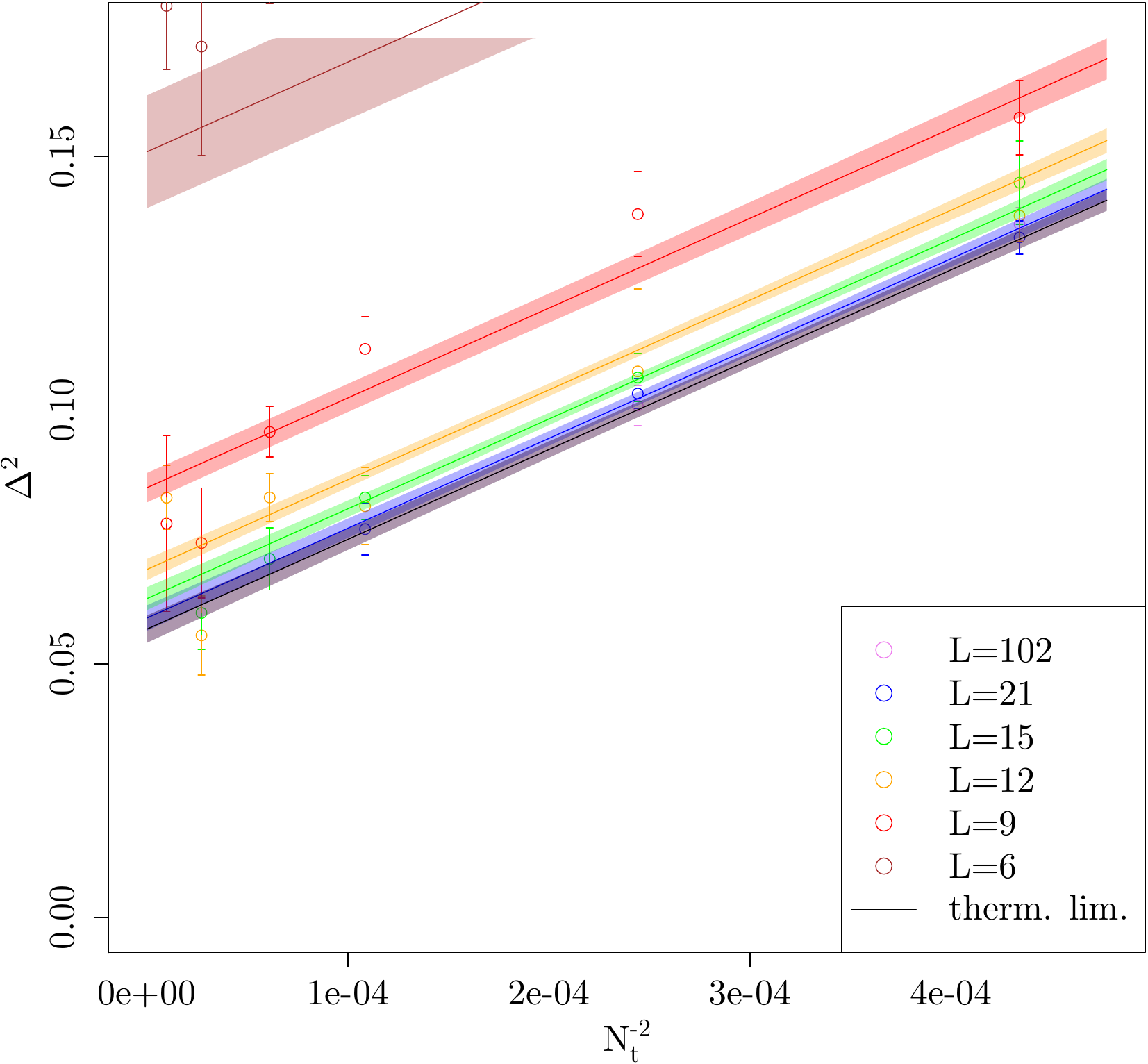}
	\caption{Simultaneous two-dimensional fit of $\Delta(N_t,L)$ (in units of $\kappa$) using Eq.~\eqref{eqn_gap_artefacts}, for $\kappa\beta=8$ and $U/\kappa=\num{3.5}$. Data points for $L<9$ have been omitted from the fit, 
		but not from the plot. Very small lattices lead to large values of $\Delta$, which are not visible on the scale of the plot.
		This fit has $\chi^2/\text{d.o.f.} \simeq \num{1.1}$, corresponding to a p-value of $\simeq \num{0.34}$.
		\label{fig:2d_gap_fit}}
\end{figure}

All our results have been extrapolated to the thermodynamic and continuous time limits in a two-dimensional fit simultaneously. In particular the single particle gap and the staggered magnetisation defined by
\begin{equation}
	m_s^2 = \frac{\sum_i S^{ii}_{--}}{V(\kappa\beta)^2}= \frac{2S^{11}_{--}+S^{33}_{--}}{V(\kappa\beta)^2},
	\label{eq:definition_m_s}
\end{equation}
have been fitted using the functional dependences
\begin{align}
	\Delta^2(L,N_t^{}) &=
	\Delta_0^2 + c_0^{\Delta} N_t^{-2}
	+ c_1^{\Delta} L^{-3}\,,\label{eqn_gap_artefacts}\\
	m_s &= m_{s,0} + c^{m_s}_0\,\delta + c^{m_s}_1 \,L^{-2}\,.\label{eqn:extrapolation_m_s}
\end{align}
An example of such a fit can be found in figure~\ref{fig:2d_gap_fit}.

\subsection{Results \label{sec:gap_results}}

\begin{figure}[ht]
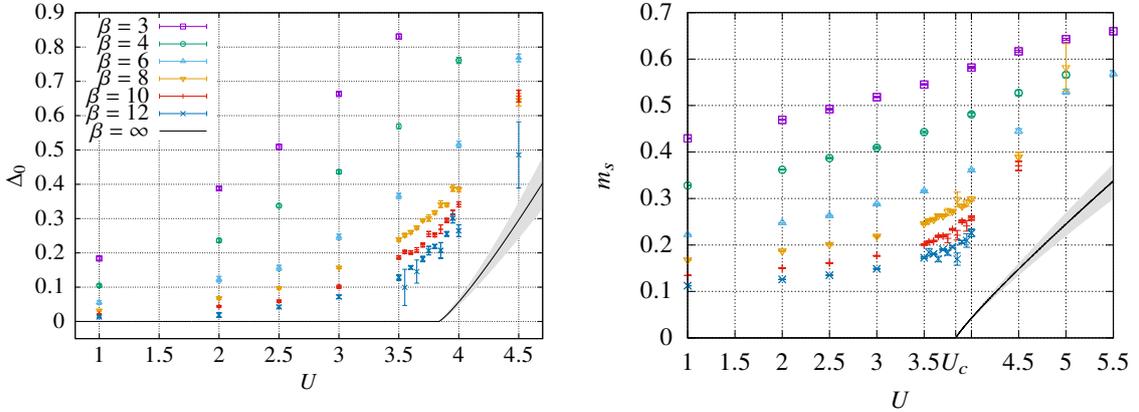

	\raisebox{0.06\height}{
		\resizebox{0.5\textwidth}{!}{{\Large
				\input{data/gap}}}}
	\resizebox{0.505\textwidth}{!}{{
			\input{data/m_s_intensive}}}
	\caption{All quantities in units of $\kappa$ and after the thermodynamic and continuum limit extrapolations. $\beta$ is the inverse temperature.
		The single-particle gap $\Delta_0(U,\beta)$ (left) and the AFMI order parameter (staggered magnetization) $m_s$ (right).  
		We also show $\Delta_0(U,\beta = \infty)$ and $m_s(U,\beta = \infty)$ as solid black lines with error band. The legend from the left plot applies to both.}
	\label{fig:both_order_parameters}
\end{figure}

Our results for $\Delta_0$ and $m_s$ are shown
in \Figref{both_order_parameters} for all values of $U/\kappa$ and $\kappa\beta$, along with an extrapolation (with error band) to zero temperature ($\beta\to\infty$).
For details as to the zero-temperature gap, see Ref.~\cite{semimetalmott}.

%

\begin{figure*}[ht]
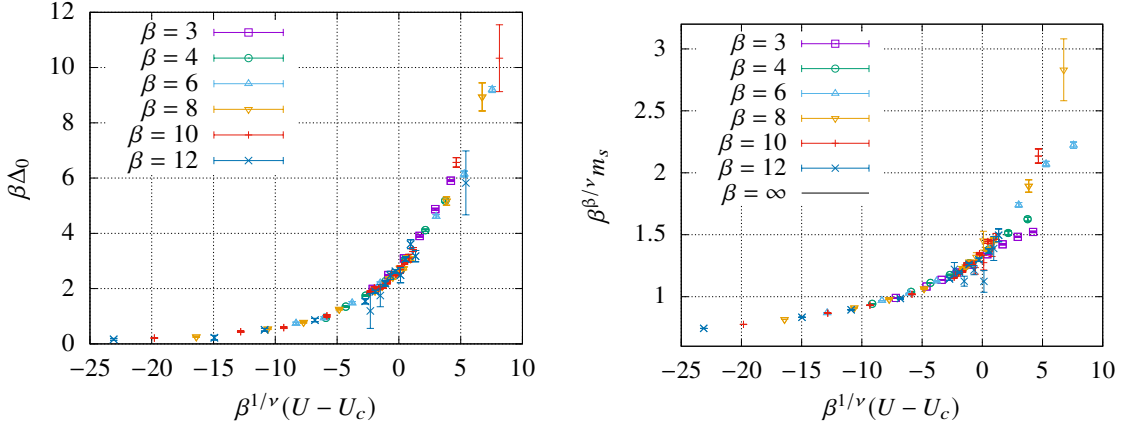

	\resizebox{0.5\textwidth}{!}{{\input{data/gap_collapse0}}}
	\resizebox{0.5\textwidth}{!}{{\input{data/m_s_collapse}}}
	\caption{Data collapse plots with the optimal parameters of $U_c$, $\nu$, and $\upbeta$ obtained from a simultaneous collapse fit to the gap $\Delta$ (left panel) and the order parameter $m_s$ (right panel).
	Note that the ``outliers'' are due to particularly small $\beta$ and are excluded from the analysis.
	 All quantities are expressed in units of $\kappa$.}
	\label{figure:structure factors}
\end{figure*}

Finally, we perform a simultaneous data collapse fit to
\begin{align}
	\beta\Delta_0 &= F(\beta^\mu(U-U_c))\,,\label{eq:collapse_delta}\\
	\beta^{\mu\upbeta} m_s &= G(\beta^\mu(U-U_c))\label{eq:collapse_m_s}
\end{align}
with some universal functions $F$ and $G$. Here $\mu=1/(z\nu)$ with the dynamical exponent $z=1$. Thus we obtain the critical parameters $U_c=\critU$, $\nu=\critNu$ and $\upbeta=\critExp$. The collapse plots are shown in figure~\ref{figure:structure factors}. The critical coupling $U_c$ and the critical exponent $\mu$ appear in both contributions~\eqref{eq:collapse_delta} and~\eqref{eq:collapse_m_s} to the collapse fit proving that gap and AFM order do indeed emerge simultaneously.

\section{Conclusions}\label{sec:conclusion}

Our work represents the first instance where the grand canonical Brower-Rebbi-Schaich (BRS) algorithm has been applied to the hexagonal Hubbard model 
(beyond mere proofs of principle), and we have found highly promising results.

We calculated the single particle gap $\Delta$ as well as all operators that contribute to the antiferromagnetic (AFM), ferromagnetic (FM), and charge density wave (CDW) order parameters of the Hubbard Model on a honeycomb lattice (all the details in Refs.~\cite{semimetalmott,more_observables}).
Furthermore we provide a comprehensive analysis of the temporal continuum, thermodynamic and zero-temperature limits for all these quantities. The favorable scaling of the HMC
enabled us to simulate lattices with $L > 100$ and to
perform a highly systematic treatment of all three limits.

The semimetal-antiferromagnetic Mott insulator (SM-AFMI) transition falls into the Gross-Neveu (GN)-Heisenberg 
$SU(2)$ universality class~\cite{PhysRevLett.97.146401}.
The GN-Heisenberg critical exponents have been studied by means of
multiple methods.
In Table~\ref{tab:crit_values}, we give an up-to-date comparison of these calculations with our results. Our value for $U_c/\kappa$ is in overall agreement with previous Monte Carlo (MC) simulations. 
For the critical exponents $\nu$ and $\upbeta$, the situation is less clear. Our results for $\nu$ (assuming $z = 1$ due to Lorentz invariance~\cite{PhysRevLett.97.146401}) agree 
best with the MC calculation (in the Blankenbecler-Sugar-Scalapino (BSS) formulation) of Ref.~\cite{Buividovich:2018yar}, followed by the FRG and large~$N$ calculations. 
On the other hand, our critical exponent $\nu$ is systematically larger than most projection Monte Carlo (PMC) calculations and first-order $4-\epsilon$ expansion results. The agreement appears to be significantly 
improved when the $4-\epsilon$ expansion is taken to higher orders, although the discrepancy between expansions for $\nu$ and $1/\nu$ persists.
Finally, our critical exponent $\upbeta$ does not agree with any results previously derived in the literature. They have been clustering in two regions. The PMC methods and first order $4-\epsilon$ expansion yielding values between $0.7$ and $0.8$, the other methods predicting values larger than 1. Our result lies within this gap at approximately $0.9$ and our uncertainties do not overlap with any of the other results.

\begin{table}
	\caption{Summary of critical couplings $U_c/\kappa$ and critical exponents $\nu$ and $\upbeta$ obtained by recent MC calculations of 
		various Hubbard models in the Gross-Neveu (GN) Heisenberg universality class, and with other methods for direct calculations of the GN Heisenberg model.
		We include brief comments of special features of each
		calculation. Note the abbreviations HMC (Hybrid Monte Carlo), AF (Auxiliary Field), 
		BSS (Blankenbecler-Sugar-Scalapino) and BRS (Brower-Rebbi-Schaich). These concepts are explained in the main text. 
		Furthermore, we denote FRG (Functional Renormalization Group).
		Our value of $\nu$~($\dag$) is given for $z = 1$~\cite{PhysRevLett.97.146401}.
		The asterisk (*) indicates that the $4-\epsilon$ exponents of Ref.~\cite{Herbut:2009vu} were used as input in the
		MC calculation of $U_c$ in Ref.~\cite{Assaad:2013xua}. Also, note the ambiguities~\cite{Otsuka:2015iba} as to the correct number of fermion components in the $4-\epsilon$ expansion
		of Ref.~\cite{Rosenstein:1993zf}.
		\label{tab:crit_values}}
	\vspace{.2cm}
	\begin{tabular}{l*{3}{S[table-format=1.7]}}
		Method & $U_c/\kappa$ & $\nu$ & $\upbeta$ \\
		\hline
		Grand canonical BRS HMC (\cite{more_observables}, present work) & 3.835(14) & 1.181(43)$^\dag$ & 0.898(37) \\
		Grand canonical BSS HMC, complex AF~\cite{Buividovich:2018yar} & 3.90(5) & 1.162 & 1.08(2) \\
		Grand canonical BSS QMC~\cite{Buividovich:2018crq} & 3.94 & 0.93 & 0.75 \\
		Projection BSS QMC~\cite{Otsuka:2015iba} & 3.85(2) & 1.02(1) & 0.76(2) \\
		Projection BSS QMC~\cite{Toldin:2014sxa} & 3.80(1) & 0.84(4) & 0.71(8) \\
		Projection BSS QMC, pinning field~\cite{Assaad:2013xua} & 3.78 & 0.882* & 0.794* \\
		GN $4-\epsilon$ expansion, 1st order~\cite{Herbut:2009vu, Otsuka:2015iba} & & 0.882* & 0.794* \\
		GN $4-\epsilon$ expansion, 1st order~\cite{Rosenstein:1993zf, Otsuka:2015iba} & & 0.851 & 0.824 \\
		GN $4-\epsilon$ expansion, 2nd order~\cite{Rosenstein:1993zf, Otsuka:2015iba} & & 1.01 & 0.995 \\
		GN $4-\epsilon$ expansion, $\nu$ 2nd order~\cite{Rosenstein:1993zf,heisenberg_gross_neveu} & & 1.08 & 1.06 \\
		GN $4-\epsilon$ expansion, $1/\nu$ 2nd order~\cite{Rosenstein:1993zf,heisenberg_gross_neveu} & & 1.20 & 1.17 \\
		GN FRG~\cite{heisenberg_gross_neveu} & & 1.31 & 1.32 \\
		GN FRG~\cite{Knorr:2017yze} & & 1.26 & \\
		GN Large $N$~\cite{Gracey:2018qba} & & 1.1823 &
	\end{tabular}
\end{table}

Thus, though we are confident to have pinned down the nature of the phase transition and to have performed a thorough analysis of the critical parameters, the values of $\nu$ and $\upbeta$ remain ambiguous. This is mostly due to a large spread of incompatible results existing in the literature prior to this work. With our values derived by an independent method we add a valuable confirmation for the critical coupling and some estimations of $\nu$ as well as a new but plausible result for $\upbeta$.

Our progress sets the stage for future high-precision calculations of additional observables of 
the Hubbard model and its extensions, as well as other Hamiltonian theories of strongly correlated electrons. 
We anticipate the continued advancement of calculations with ever increasing system sizes, through the leveraging of additional state-of-the-art techniques from lattice QCD, 
such as multigrid solvers on GPU-accelerated architectures. We are actively pursuing research along these lines.


\section*{Acknowledgements}

We thank Jan-Lukas Wynen for helpful discussions on the Hubbard model and software issues. We also thank Michael Kajan for proof reading and for providing a lot of detailed comments.
This work was funded, in part, through financial support from the Deutsche Forschungsgemeinschaft (Sino-German CRC 110 and SFB TRR-55).
During this work E.B. was supported by the U.S. Department of Energy under Contract No. DE-FG02-93ER-40762.
The authors gratefully acknowledge the computing time granted through JARA-HPC on the supercomputer JURECA~\cite{jureca} at Forschungszentrum J\"ulich.
We also gratefully acknowledge time on DEEP~\cite{DEEP}, an experimental modular supercomputer at the J\"ulich Supercomputing Centre.
The analysis was mostly done in \texttt{R}~\cite{R_language} using \texttt{hadron}~\cite{hadron}.
We are indebted to Bartosz Kostrzewa for helping us detect a compiler bug and for lending us his expertise on HPC hardware.


\FloatBarrier
\bibliographystyle{JHEP}
\bibliography{cns}

\end{document}